\documentclass[prl,aps,twocolumn,floats,superscriptaddress,longbibliography]{revtex4-1}
\include{header}
\usepackage{soul}
\usepackage{manfnt}
\usepackage{xcolor}
\usepackage{multirow}
\usepackage{amsmath}
\usepackage{graphicx}
\graphicspath{{figs/}}
\usepackage{amssymb}
\usepackage{epsfig}
\usepackage[utf8]{inputenc}
\usepackage[T1]{fontenc}
\usepackage{newtxtext,newtxmath}
\usepackage{pdfrender}
\newcommand*{\boldgreek}[1]{%
  \textpdfrender{%
    TextRenderingMode=FillStroke,%
    LineWidth=.35pt,%
  }{#1}%
}
\usepackage{simplewick}
\usepackage[colorlinks = true]{hyperref}
\newcommand{\cube}{{\square}}
\newlength{\upit}\upit=0.1truein

\newcommand{\ltappr}{{{\lower4pt\hbox{$<$} } \atop \widetilde{ \ \ \ }}}
\newlength{\bxwidth}\bxwidth=1.5 truein

\newcommand{\dg}{^{\dagger }}

\newcommand{\gtappr}{{{\lower4pt\hbox{$>$} } \atop \widetilde{ \ \ \ }}}

\newcommand{\bk}{{\bf{k}}}
\newcommand{\bx}{{\bf{x}}}
\newcommand{\by}{{\bf y}}
\newcommand{\ba}{{\bf{a}}}
\newcommand{\bR}{{\bf{R}}}
\newcommand{\pmat}[1]{\begin{pmatrix}#1\end{pmatrix}}

\newcommand{\be}{\begin{equation}}
\newcommand{\ben}{\begin{equation*}}
\newcommand{\ee}{\end{equation}}
\newcommand{\bea}{\begin{eqnarray}}
\newcommand{\eea}{\end{eqnarray}}

\newcommand{\een}{\end{equation*}}

\newcommand{\bmx}{\begin{array}}
\newcommand{\emx}{\end{array}}
\newcommand{\bean}{\begin{eqnarray*}}
\newcommand{\eean}{\end{eqnarray*}}

\newcommand{\p}{\partial}
\newcommand{\s}{\sigma}

\newcommand{\rd}{\mbox{d}}

\newcommand\ltdash{\raise-0.7pt\hbox{$\scriptscriptstyle |$}}

\newlength{\figwidth}
\newlength{\shift}
\newcommand{\fgb}[3]
{\begin{figure}[b]\vspace*{-0cm}\centerline{\includegraphics[width=\figwidth]{#1}}\vskip
-0.2cm \caption{#3}\label{#2}\end{figure}}
\begin{document}
\title{A solvable 3D Kondo lattice exhibiting pair density wave, odd-frequency pairing
and order fractionalization}
% Alexei -  with the alphabetical order of authors, we have the catchy
% acronym of CPT - our model becomes the CPT model.
% If this is OK with you. 
\author{Piers Coleman}
\affiliation{
Center for Materials Theory, Department of Physics and Astronomy,
Rutgers University, 136 Frelinghuysen Rd., Piscataway, NJ 08854-8019, USA}
\affiliation{Department of Physics, Royal Holloway, University
of London, Egham, Surrey TW20 0EX, UK.}
\author{Aaditya Panigrahi}
\affiliation{
Center for Materials Theory, Department of Physics and Astronomy,
Rutgers University, 136 Frelinghuysen Rd., Piscataway, NJ 08854-8019, USA}
\author{Alexei Tsvelik}
\affiliation{Division of Condensed Matter Physics and Materials
Science, Brookhaven National Laboratory, Upton, NY 11973-5000, USA}
\date{\today}
\pacs{PACS TODO}
\begin{abstract}
{  
%We describe a mechanism for order fractionalization in a
%two-dimensional Kondo lattice model, in which electrons interact with
%a gapless spin liquid of Majorana fermions described by the
%Yao-Lee (YL) model.  
%When the Kondo coupling to the conduction electrons exceeds a
%critical value,  the model develops a superconducting
%instability into a state with a a spinor order parameter with 
%charge $e$ and spin $S=1/2$. The broken symmetry state develops a
%gapless Majorana Dirac cone in the bulk.
%By including an appropriate gauge string, we can 
%show that the charge $e$, spinorial order develops 
%off-diagonal long range order that allows electrons to coherently tunnel
%arbitrarily long distances through the spin liquid. 
The Kondo
lattice model
plays a key role in our
understanding of quantum materials, 
but a lack of small parameters has posed a
long-standing problem. We 
present a 3 dimensional 
S$=1/2$ Kondo lattice model  describing  a  spin liquid within  an
electron sea. Strong correlations in the spin liquid 
are  treated exactly, enabling a controlled analytical approach. 
Like a Peierls or BCS phase, a logarithmically divergent 
susceptibility leads to an instability into a new phase
at arbitrarily small Kondo coupling. Our solution captures 
a plethora of emergent phenomena, including odd-frequency pairing, pair density wave
formation and order fractionalization. The 
ground-state state  is a pair
density wave with a fractionalized charge $e$,
$S=1/2$ order parameter, formed between electrons and
Majorana fermions. 
}
\end{abstract}

\maketitle
%\item Introduction. Fractionalized order in a 3D Kondo lattice.
%The wonders of KL and how they are relevant today.

 The rich physics of the Kondo lattice,  describing  an array of
 local moments interacting with 
an electron sea, plays a key role in our
understanding of quantum materials, from
heavy fermion compounds to twisted moir\' e lattices \cite{wirth2016,aline21,potter21,bernevig21}.   A key element of this
model is the fractionalization 
of spins into heavy fermions, producing a large Fermi surface. These
interactions have to date been treated with approximate methods,
such as the large-N expansion\cite{read1983,Coleman1983,auerbach,millis87,ColemanPRL87} and
dynamical mean field theory\cite{Si_Nat2001}. 

Our work builds on 
a series  of important developments in the
theory of Kitaev models
and their connection with the Kondo lattice\cite{Hermanns,Hermanns2,seifert2018,choi2018,Freire21,miranda20}. 
Earlier variants of Kitaev-Kondo lattices
include models that couple the original, spin-gapped
Kitaev spin liquid to a conduction
sea
\cite{seifert2018,choi2018,Freire21},
and models that couple  
two dimensional Yao-Lee spin liquid to a conduction sea via an
octupolar coupling\cite{miranda20}.

%A lack of small parameters in this model  has posed a long-standing
%obstacle to its accurate theoretical treatment
Here we 
introduce a 3-dimensional 
Kondo lattice model which couples a  $Z_{2}$ spin
liquid with a Fermi surface to a conduction sea, in which the Kondo lattice
physics can be treated analytically to leading logarithmic accuracy. 
The  spin fluid is a three dimensional generalization of the
Yao Lee spin liquid\cite{yaolee11}, { embedded on a hyper-octagonal
lattice\cite{Hermanns}, chosen
because its cubic, trivalent structure
gives rise to an exactly solvable, gapless spin liquid whose
gapless  Majorana excitations lie on a Fermi surface.}  Like 
the Kitaev spin liquid, the $Z_{2}$ gauge fields associated
with the fractionalized spins are static and can be treated
exactly \cite{yaolee11,kitaev}. 

%Order Fractionalization hypothesis. 
Recent work has hypothesized that  hybridization between
electrons and fractionalized excitations can give rise to a new kind
of {\sl fractionalized order} 
with half-integer  quantum numbers\cite{komijani2019order,tsvelik22}. Our Kondo
lattice model provides a rigorous example of this phenomenon. 
In particular,  at half filling, { the judicial choice of the lattice guarantees a perfect}  nesting between spinon and  electron 
Fermi surfaces and allows us to sum the leading logarithmic divergences, 
establishing an instability at infinitesimal Kondo coupling
into a pair-density wave
with a charge $e$, $S=1/2$ order
parameter, which induces odd-frequency pairing amongst the conduction
electrons and also gives rise to a neutral, Majorana Fermi
surface. 
\begin{figure}[h]
	\begin{center}
	\includegraphics[scale=0.15]{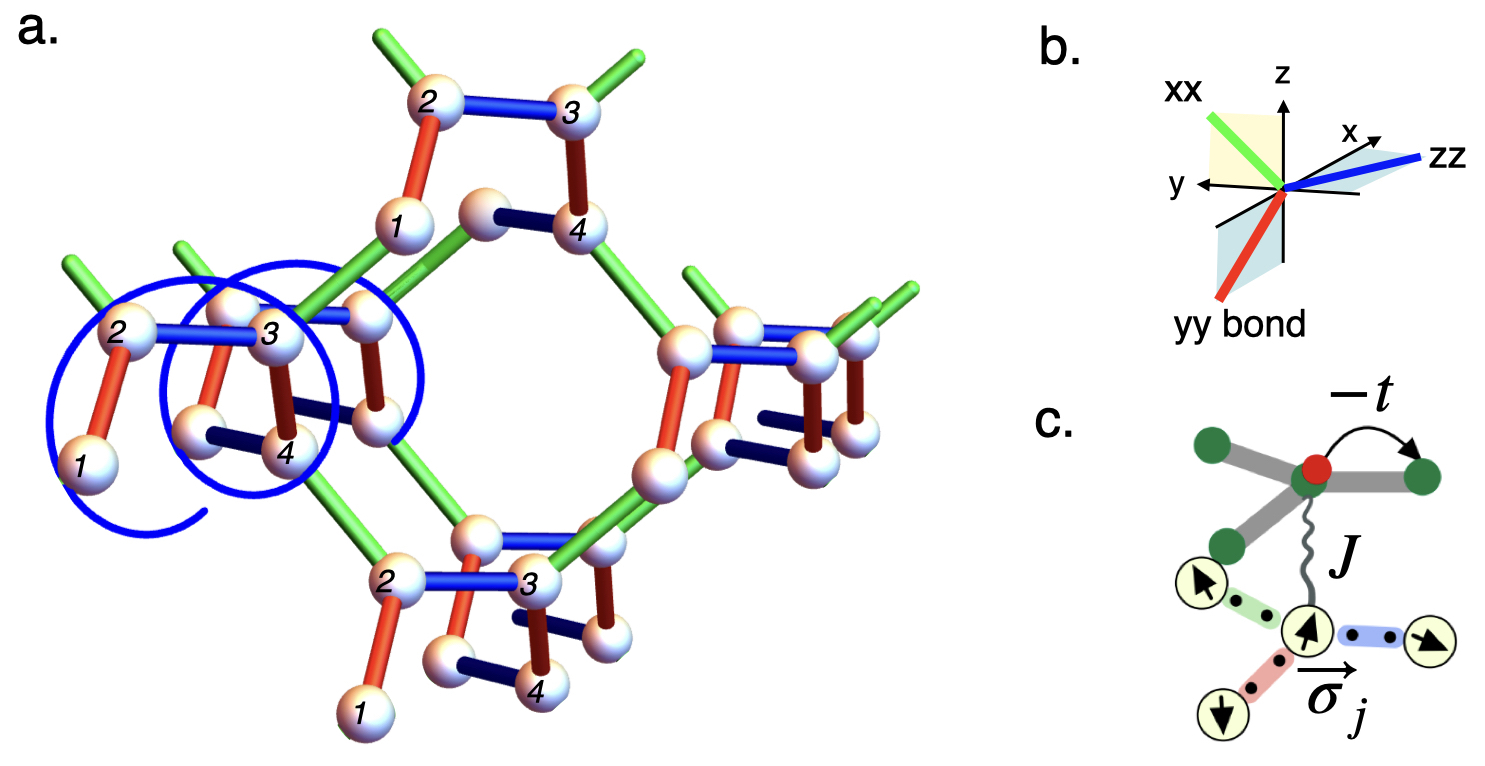}
	\caption{(a) Hyper-octagonal lattice: 
a  four-atom 
coil ($ 1,2,3,4$) on a BCC lattice gives rise to alternating
square and
	octagonal spirals\cite{Hermanns}.(b) Anisotropic coupling of
	orbital degrees of freedom according to the plane in which the
	bond lies. (c) Kondo coupling between the Yao-Lee
	spin liquid and the electron sea.}\label{fig1}
	\end{center}
\end{figure}
%\begin{itemize}
%\vskip -0.3in

Our model Hamiltonian $H=H_{C}+H_{YL}+H_{K}$, where
\begin{eqnarray}
	H_{C}&=&-t\sum_{<ij>}(c^{\dagger}_{i\sigma}c_{j\sigma}+{\rm
	H.c})- \mu \sum_{j}c\dg_{j\sigma }c_{j\sigma },\label{HC}\cr
H_{YL }&=&( {K}/{2})\sum_{<ij>}
(\vec{\sigma }_{i}\cdot\vec{\sigma }_j)\lambda^{\alpha_{ij}}_i\lambda^{\alpha_{ij}}_j,\label{YLm}\cr
H_{K}&=&
J\sum_{j}
\vec{S}_j\cdot ( c\dg_{j}\vec{\sigma }c_{j}).
\end{eqnarray}
Here $\langle i,j\rangle $ are neighboring sites on the
hyper-octagonal lattice\cite{Hermanns}, a  trivalent body centered cubic (BCC) crystal 
with four  atoms per primitive unit cell, coiled around a helix
to form alternating square and octagonal
spirals(Fig. \ref{fig1}).
$H_{C}$ describes hopping electrons 
while $H_{YL}$
describes a three-dimensional
Yao-Lee (3DYL) spin liquid on the same lattice, 
with an orbital and spin
degree of freedom at each site, denoted
by Pauli 
operators $\lambda^{a}_{j}$ ($a=1,2,3$)  and spins
$\vec{ S}_{j}=\vec \sigma _{j}/2$,  respectively.
The $\alpha_{ij}=x,y,z$ label  anisotropic $xx$, $yy$ and $zz$ orbital
interactions on bonds that lie in the $yz, zx$ and $xy$ planes
respectively. Finally, 
$J$ is an  antiferromagnetic 
Kondo coupling between the electrons and local moments.

%\item  The model - Yao Lee + Kondo.  Talk about how the gapped visons
%remove gauge fluctuations, so that fractionalized particles become
%well-defined at low temperatures.  

{\it Fermionization}.   The Yao-Lee model \cite{yaolee11} belongs to a
family of $Z_2$ Kitaev spin liquids which can be solved  exactly using
fermionization. 
Following \cite{yaolee11} we represent spin and orbital operators as products of Majorana fermions 
 	$\vec{\sigma}_j=-i\vec{\chi}_j\times \vec{\chi}_j$ and 
$	\vec{\lambda}_j=-i\vec{b}_j\times \vec{b}_j
$, where we use the normalization 
$\{\chi_{i}^{a},\chi_{j}^{b} \}= \delta_{ij}\delta^{ab}$ .
The spin-orbital operator is given by 
$\vec{\sigma}_{j}\lambda_j^{\alpha } = -2i D_{j}\vec{\chi}_{j}b_{j}^{\alpha }$
where the constants $D_{j}= 8i \chi^{1}_{j}\chi^{2}_{j}\chi^{3}_{j}
b^{1}_{j}b^{2}_{j}b^{3}_{j}= \pm 1$ commute with $H$. In the gauge
$D_{j}=1$, 
\begin{equation}\label{}
H_{YL} ={K}\sum_{<i,j>} u_{ij} (i \vec{\chi }_{i}\cdot \vec \chi_{j})
\end{equation}
where $u_{ij}= - 2i b_{i}^{\alpha_{ij}}b_{j}^{\alpha_{ij}}= \pm 1$ is
a $Z_{2}$ gauge field that commutes with the Hamiltonian.  

The 3DYL model  describes free fermions moving in a
static $Z_2$ gauge field.  This model shares many of the properties of a 2D Kitaev
spin liquid, most notably, the presence of gapped $Z_{2}$ flux excitations. 
%: in particular the gauge fields $u_{ij}$ are
%static constants of the motion.  
On the hyper-octagonal
lattice these are described by Wilson loops - products of
the gauge fields $W= \prod u_{(i,j)}= \pm 1  $ around closed ten and
twelve-fold loops (where $(i,j)$ orders the sites $i$ and $j$ 
along  $xx,\ yy$ and $zz$ bonds so that
the site furthest in the $y$, $z$ and $x$
directions respectively, is placed first \cite{Hermanns}). In the spin liquid ground-state,
all loops are trivial $W=1$\cite{Hermanns}; flipping the sign of a
Wilson loop creates a flux excitation (vison), with an energy
determined as a fraction of $K$.  Unlike 2D Kitaev spin liquids,
the 3 dimensional models
models undergo an Ising phase transition 
into a Higgs phase where gauge fluctuations are suppressed and Wilson loops
develop long-range order 
\cite{janssen20,mishenko17,transition}. 
For the Kitaev model on
the hyper-octagon lattice the { transition at}
$T_{c1}\sim 0.012K $\cite{mishenko17,transition} gives rise to a spin-gap, but
in the 3DYL, spin excitations occur without the creation of visons, 
eliminating the spin-gap; moreover,
the three Majorana modes enhance $T_{c1}$ by a factor of three.
Below $T_{c1}$
gauge fluctuations are suppressed and
the Majorana fields describe coherent, fractionalized
spin excitations.

%\begin{itemize}

\begin{figure}[b]
	\begin{center}
	\includegraphics[scale=0.15]{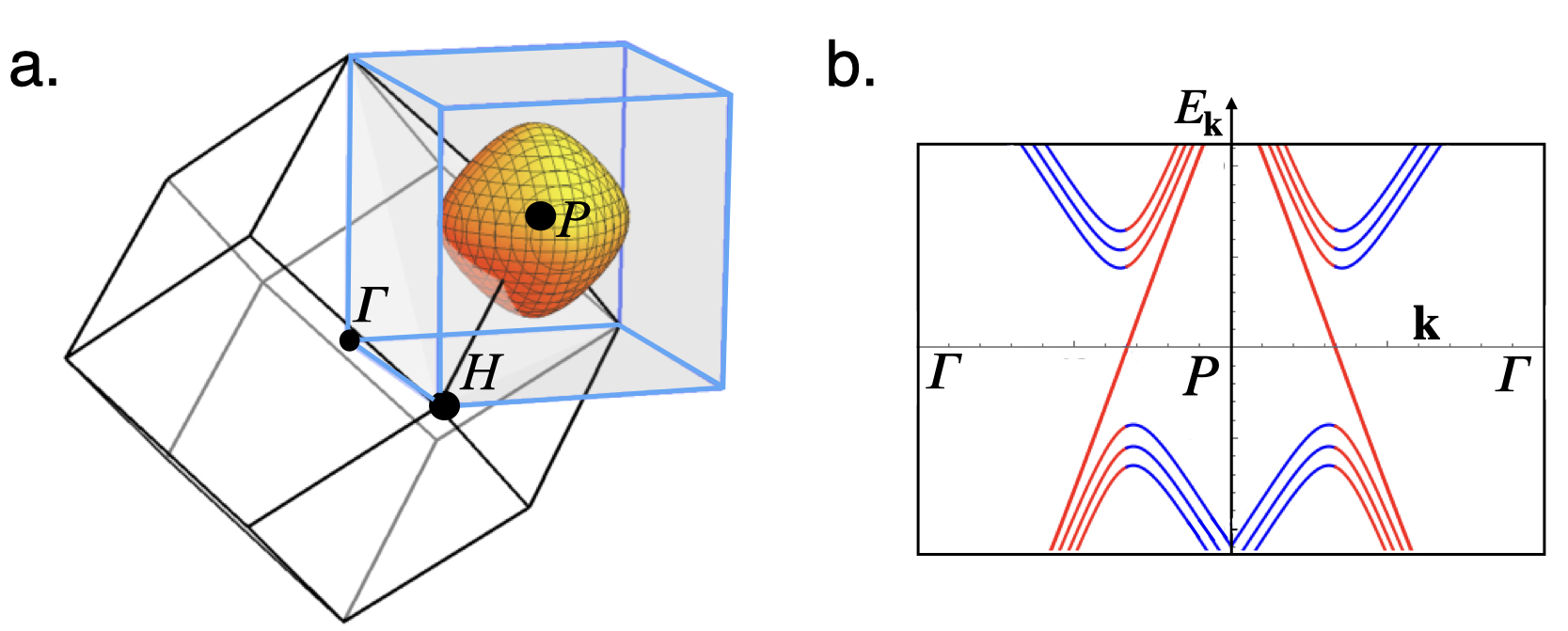}
	\caption{(a) Majorana Brillouin zone  of the  hyper-octagonal
	lattice showing coincident conduction and Majorana Fermi surfaces
	around the P point. (b) Hybridization of Majorana modes (in
	blue) with conduction band (in red) leaves one
conduction Majorana band decoupled, forming a neutral Majorana 
surface. 
}\label{fig2}
	\end{center}
\end{figure}

In the ground-state, choosing a gauge where
$u_{(i,j)}=1$ and taking into account that  $\chi_{-\bk} = \chi\dg_{\bk}$, 
%the corresponding  Hamiltonian is %of normal state $H = H_{c}+H_{YL}$, where 
\begin{equation}\label{YLm2}
H_{YL}=  K\sum_{\bk \in \cube}
\vec{ \chi}\ \dg _{\bk \alpha }
h (\bk )_{\alpha \beta }
\vec{\chi}_{\bk \beta },
\end{equation}
where $\alpha, \beta  \in [1,4]$ are site indices, while 
\begin{equation}\label{}
h (\bk )=
\pmat{
0 & i &  ie^{-i\bk\cdot \ba_2} & ie^{-i\bk\cdot \ba_1 }
\\-i & 0 & -i & ie^{-i \bk\cdot\ba_3}
\\- ie^{i\bk\cdot \ba_2} & i & 0 & -i
\\-ie^{i\bk\cdot \ba_1 } &  -ie^{i \bk\cdot\ba_3} & i & 0
},
\end{equation}
and 
%\begin{equation}
$	\ba_1=(1,0,0);\ba_2=\frac{1}{2}(1,1,-1);\ba_3=\frac{1}{2}(1,1,1),$
	%\label{eqn: BravaisVec}
%\end{equation}
are the primitive BCC lattice vectors. 
Since $\chi_{-\bk} = \chi\dg_{\bk}$, the momentum sum is restricted to
half
the Brillouin zone, corresponding to a cube 
($\cube$) of side length $2\pi$ centered at the
$P$ point at  $(\pi,\pi,\pi)$.
The spectrum $E_{\bk }\equiv K \epsilon (\bk )$,
determined by ${\rm det} [\epsilon \underline{1}-h ({\bk })]=0$, or
\begin{equation}\label{eqn: characteristic}
\epsilon^4-6\epsilon^2 -8\epsilon (s_{x}s_{y}s_{z})
+ [9- 4 (s_{x}^{2}+s_{y}^{2} + s_{z}^{2})]=0,
\end{equation}
(where $s_{l}\equiv \sin (k_{l}/2)$, $l=x,y,z$),
contains a {\sl single } Fermi surface centered at $P$\cite{Hermanns2}
 (Fig. \ref{fig2}). 

%\begin{figure}[h]
%\begin{subfigure}[a]{0.3\textwidth}
%			\includegraphics[scale=0.2]{3DKK_Aaditya_Notes/BZ3}
%	\end{subfigure}
%	\begin{subfigure}[b]{0.3\textwidth}
%			\includegraphics[scale=0.2]{3DKK_Aaditya_Notes/BandStructureGapless.png}
%	\end{subfigure}

%	\caption{(a) the Brillouin zone of  hyper-octagon lattice; the blue lines show  the directions along which the band structure for the model is plotted. (b)The band structure for the Majorana fermion in the Kitaev model on hyper-octagon lattice at $K=\frac{1}{3}$ along the line joining high-symmetry points $\Gamma\rightarrow P\rightarrow N \rightarrow H \rightarrow \Gamma \rightarrow N \rightarrow P 
%	\rightarrow H$ }
%	\label{pic: BandStructure2}
%	\end{center}
%\end{figure}
Since the electrons and majoranas 
move on the same lattice, at half filling their Fermi surfaces 
are perfectly nested and can be brought into coincidence 
by applying  a gauge transformation to the electrons, 
\begin{eqnarray}
	&& (c_{1},c_{2},c_{3},c_{4})_{\vec{R}}\rightarrow e^{{-i} {\bf
	Q\cdot R}} (c_{1},ic_{2},c_{3},-ic_{4})_{\vec{R}}, \label{Qvector}%e^{-i\pi(n_1-\frac{1}{2}(n_2-n_3))
\end{eqnarray}
where $	{\bf Q} = { (\pi, \pi,\pi )}$ and $ ~~ {\bf R} = n_1{\bf a}_1
+ n_2{\bf a}_2 + n_3{\bf a}_3 $ locates the unit cell.
In this gauge, 
\begin{equation}\label{HC2}
H_{c}=  \sum_{\bk\in BZ }
c\dg _{\bk \sigma \alpha } [-t\  h (\bk) -\mu\underline{1}]_{\alpha \beta }
c_{\bk \sigma \beta },
\end{equation}
and $H_{YL}$(\ref{YLm2}) have the same
form. 
%In $H_{c}$ the momentum sum is over the full Brillouin zone (BZ).
At low  temperatures, where flux excitations can be ignored, 
we can rewrite the Kondo interaction in terms of the spin
$\vec{S}_{j}= - (i/2) \vec{\chi }_{j}\times \vec{\chi }_{j}$ and 
decouple it using a Hubbard-Stratonovich
transformation, in terms of a charge $e$ spinor,
$V_{j}=(V_{j\uparrow}, V_{j\downarrow})^{T}$,
 \begin{eqnarray}
H_{K} 
=
\sum_{j}\left[(c\dg_{j}\vec{\sigma }V_{j})\cdot \vec{\chi
}_{j}+
{\rm H.c.}\right]+2\frac{V^{\dagger}_jV_j}{J} \label{decouple}
\end{eqnarray}
The terms multiplying $\vec\chi_{j}$ must themselves be Majorana
fermions, enabling us to  rewrite $H_{K}$ in the compact form 
\begin{equation}\label{}
H_{K}= \sum_{j}\left[ 
 -i {\rm V_{j}} (\vec{c}_{j}\cdot \vec{\chi }_{j}) + \frac{{\rm V}_{j}^{2}}{J}\right],
\end{equation}
where we have cast $V_{j}= ({\rm V}_{j}/\sqrt{2}) (z_{\uparrow
j}, z_{\downarrow j})^{T}$ in terms of a normalized spinor and a real
amplitude
$\rm V_{j}/\sqrt{2}$,   dividing
the
electrons into four Majorana components, $(c_{j}^{0}, \vec{c}_{j})$,
\begin{equation}\label{}
\pmat{c_{j\uparrow} \cr c_{j\downarrow }}
= \frac{1}{\sqrt{2}} (c^{0}_{j}+ i \vec{c}_{j}\cdot \vec{\sigma })\
\pmat{z_{j\uparrow}\cr  z_{j\downarrow } }.
\end{equation}
Thus in a coherent Kondo lattice, the vector components of the electron
hybridize with the spinons and the scalar component decouples.  

Now the field $V_{j}$ is a fluctuating field inside the path integral,
but the nesting between
the Majorana and conduction Fermi surfaces ensures that it has a
susceptibility to condense that is logarithmically divergent in temperature or
chemical potential. Like a Cooper or  Peierls instability,
this ensures that instability into a condensed phase 
must occur for arbitrarily weak Kondo coupling.

We shall now focus on the stable, uniform condensate 
 $V_{j}= V$.  
At half-filling ($\mu=0$), vector and scalar electron components 
decouple, so that 
 %given by $H= H_0 + H_{MF}$, where $H_0$ is the noninteracting Hamiltonian (\ref{HC3}) for the band Majorana mode $\Psi_0$ and $H_{MF}$ describes the hybridization of three band and three spin liquid Majorana fermions: 
\begin{eqnarray}
H = \sum_{{\bk}\in\cube}\left[-tc_\bk^{0\dagger}h_{\bk }c^{0}_{\bk }  +
\vec{\psi}^{\ \dagger}_{\bk }\pmat{
	-t h_\bk &-i{\rm V}
	\cr i{\rm V} & K\hat h_\bk }
\vec{\psi}_{\bk }\right]
+\frac{N{\rm V}^{2}}{J},\cr
\end{eqnarray}  
where $\vec\psi_{\bk } = ( \vec c_{\bk },\vec\chi_{\bk
})^{T}$, $N$ is the number of sites and we denote $
h_{\bk }\equiv h (\bk )$.

As we now demonstrate, at a temperature $T_{c2}$, below the Ising
phase transition $T_{c1}$, the system
undergoes a second phase transition where the spinor order parameter $V$
condenses (Fig.~\ref{Fig3}a). 
 To demonstrate the instability, we note that since only states close to the Fermi
 surface contribute at small $J << K, t$, we can project the
 Hamiltonian onto the band with a Fermi surface,  with a dispersion
 for the band and the spin liquid fermions being equal to $-t\epsilon(k)$ and
 $K\epsilon(k)$ respectively. The calculations then can be done analytically;
 we will restrict ourselves to the simplest case $\mu =0$. Summing the
 leading logarithmic ladder diagrams, we find the
critical temperature $T_{c2}$ is defined by the condition $ J \chi_{1e}
 (T_{c2})=1$, where
\begin{equation}
\chi_{1e}(T)= \frac{1}{2}
 \int \frac{d^3 k}{(2\pi)^3} 
\frac{\tanh[\frac{\beta t\epsilon({\bf k})}{2}] + 
\tanh[\frac{\beta K\epsilon({\bf k})}{2}]}{2({K+t})\epsilon (\bk )}
,\end{equation}
is the charge $e$ pairing  susceptibility\cite{SM}. 
To logarithmic accuracy, $\chi_{1e} (T)=  
\frac{\rho}{1+K/t} \ln  [W/T]$, 
where $\rho =2\sqrt 3 /\pi^2 t$ is the conduction density of
states,
giving $T_{c2} = W \exp\Big(- \frac{1+K/t}{\rho J}\Big)$, so a phase transition into the order
fractionalized state will occur for {\sl  arbitrarily small} Kondo coupling.
Of course, deviations from particle-hole symmetry at finite
chemical potential destroy the nesting, so when $\mu\neq  0$, a
transition takes place at finite $J>J_{C}$ from an FL$^{*}$ state\cite{Senthil_PRL2003}
with a small Fermi surface,
 into the order-fractionalized state
(Fig.~\ref{Fig3}a).
The important point however, is that in the vicinity of particle-hole
symmetry, the broken symmetry state is rigorously established. 

 Below $T_{c2}$ the vector Majorana modes are gapped, leaving behind
a single, coherent  Majorana mode $c^{0}_{\bk }$ of the conduction
 electrons. This feature is robust and is related to the mismatch between the quantum numbers of the itinerant Dirac fermions  and Majorana triplet of the Yao-Lee spin liquid. The spectrum of the gapped fermions close to the Fermi surface is given by 
 \begin{equation}
 E_{\pm} = (K-t)\epsilon({\bk})/2 \pm \sqrt{(K+t)^2\epsilon^2({\bf k})/4 + |V|^2}.
 \end{equation}

To understand the nature of the fractionalized order, it is useful to 
consider the fractionalized order parameter $\hat v (\bx_{j})= - J (\vec{\sigma}
\cdot \vec{\chi }_{j})c_{j}$. This quantity carries a $Z_{2}$ charge,
and by Elitzur's theorem, can not develop long range order.
On the other hand, we know that the $\chi_{j}$ field represents a physical
degree of freedom at low temperatures, where $Z_{2}$ fluctuations have
become massive. To reconcile this situation, we must consider the
gauge invariant density matrix 
\begin{equation}
\rho (x,y)=\langle \hat v(\bx )W (\bx ,\by) \hat v^{\dagger}(\by)\rangle 
\rangle \xrightarrow[]{|\bx-\by|\rightarrow \infty} V (\bx )V\dg (\by)
\end{equation}
where $W (\bx ,\by)= \prod u_{(l+1,l)}$ is a Wilson line connecting
the sites $\bx, \by$.  Once $T<T_{c1}$ the Wilson lines 
are not only constants
of motion, but they are independent of the path between the two sites
$\bx $ and $\by $.  
We can calculate the gauge invariant quantity in the gauge where 
$u_{(i,j)}=1$ so that $W=+1$, and in this way, 
we can be sure that the gauge invariant density
matrix asymptotically factorizes into a product of well-defined spinor
order\cite{tsvelik22}. 
In short, once $T< T_{c1}$ where the absence of visons guarantees that typical 
Wilson lines are  equal to +1, 
this fractionalized long-range order is guaranteed to develop. 

\figwidth=1.\columnwidth
\fgb{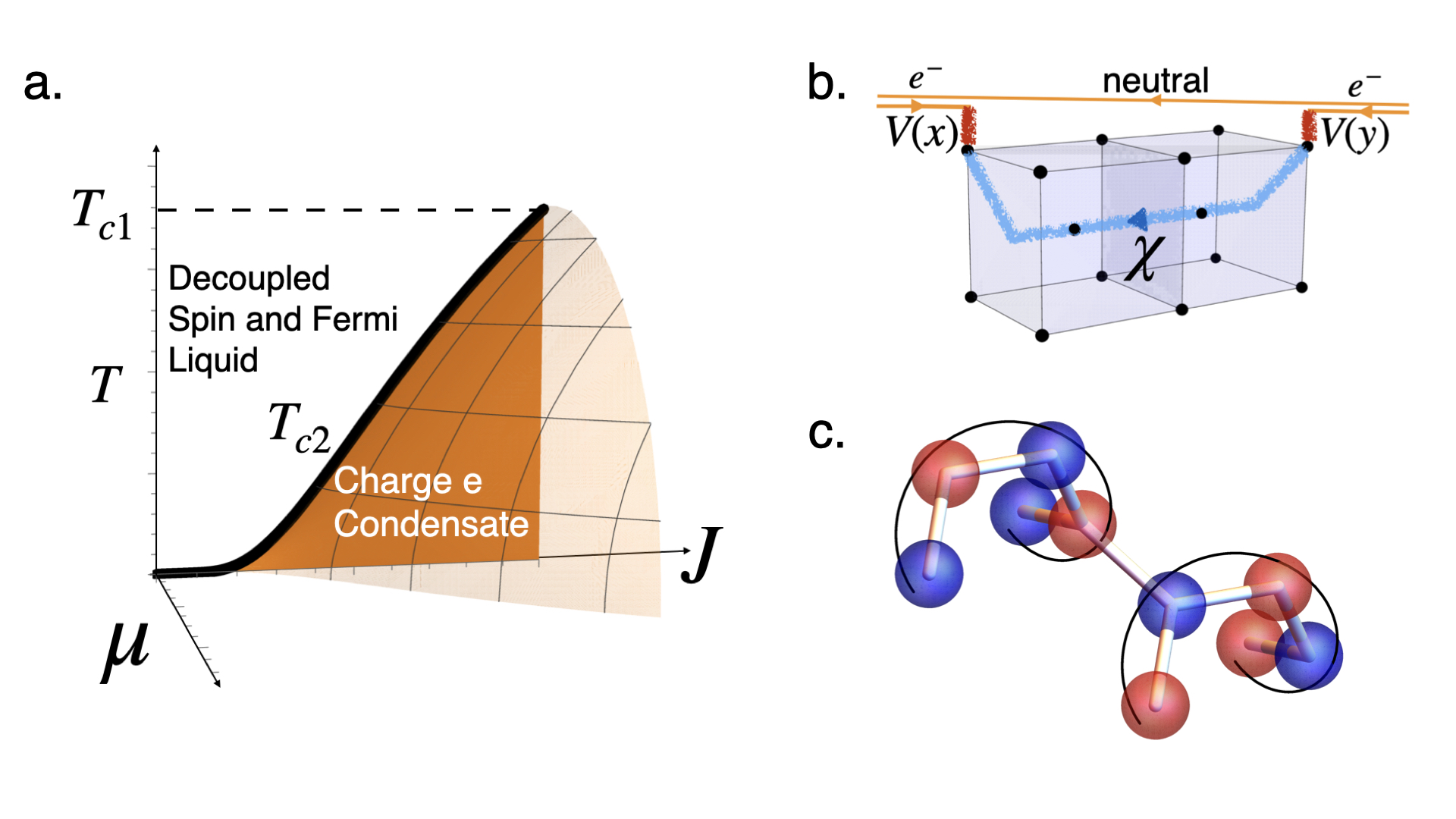}{Fig3}{a) Schematic Phase
diagram for the 3D CPT model: below  $T_{c1}$, the Majorana fermions in the
spin liquid become phase coherent and at $T_{c2}$ the charge $e$
condensate develops. At particle-hole symmetry ($\mu=0$) 
the charge e condensate forms for arbitrarily small Kondo coupling. b)
Development of coherent charge $e$ condensate allows tunneling through
the spin liquid over arbitrarily long distances, liberating a
coherent, quasi-neutral conduction mode. {c)Staggered configuration of 
the gap function $\Delta(R) \sim e^{i 2 {\bf Q}\cdot {\bR}}(\hat{\bf
d}^1+i \hat{\bf d}^2)\cdot \vec{\sigma} (1,-1,1,-1)$ at the four sites
of the unit cell, forming a pair-density wave. }} 

One of the physical manifestations of this fractionalized ODLRO, is
the development of long-range tunneling of the electrons through the
spin liquid, which manifests through the development of odd-frequency
triplet pairing and the emergence of a Majorana Fermi surface (Fig \ref{Fig3}b).  
The self-energy for the
electrons that describes the coherent tunneling through the spin liquid is
given by 
\begin{eqnarray}\label{}
\Sigma (\bk ,\omega)&=&
(1-{\cal  Z}{{\cal Z}\dg })\Sigma_{0} (\bk ,\omega), \cr
\Sigma_{0} (\bk ,\omega)&=& \frac{{\rm V}^{2}}{\omega -  K h (\bk )}.
\end{eqnarray}
Here ${\cal  Z} = \frac{1}{\sqrt{2}} (z_{\uparrow},z_{\downarrow}, -
z_{\downarrow }^{*}, z_{\uparrow}^{*})^{T}$ is a Ballian Werthammer
spinor, and the combination $P={\cal Z}{\cal Z}\dg $ 
projects out the unhybridized scalar component of the conduction sea.
Without the projector, this scattering would describe the resonant
scattering of electrons in a Kondo
insulator, but the elimination of the scalar
component means that a 
tunneling  electron that re-emerges into the spin liquid
loses all knowledge of its original charge, allowing it to emerge as
a  hole, producing resonant Andreev scattering. On the Fermi surface,
where $h (\bk )$ has vanishing eigenvalues
$\Sigma (\bk ,\omega)\sim 1/\omega$  describes odd frequency pairing that
is infinitely retarded in time. 
One of the clearest manifestations of 
long-range order in the charge $e$ order, is that the  
projective nature of the scattering self-energy remains coherent 
in momentum space, 
allowing the residual scalar Majorana conduction electron $c_{0}$
to propagate coherently over arbitrarily long distances. 

To examine this resonant  pairing in more
detail, it is useful to construct the composite order parameter, formed
from bilinears of the $z$ field, 
\begin{equation}\label{}
\hat {\bf d}^{1}_{j}+i \hat {\bf d}^{2}_{j} 
=
z^{T}_{j}(-i\sigma_{2})\mathbf{\sigma }z_{j}, \ \hat  {\bf
d}^{3}_{j} = z\dg_{j}\mathbf{\sigma }z_{j}.
\end{equation}
The triad $(\hat {\bf d}^{1}_{j},\hat
{\bf d}^{2}_{j},\hat {\bf d}^{3}_{j})$ describes co-existing magnetic
and superconducting order. 
We can then divide the self-energy 
into normal and pairing components 
\begin{equation}\label{}
\Sigma = \Sigma_{N}+\Delta (\bk ,\omega)\tau_{+}+ \Delta \dg (\bk
,\omega)\tau_{-},
\end{equation}
where
\begin{eqnarray}\label{comps}
\Sigma_{N} (\bk ,\omega)&=&
\tfrac{1}{4}
\left(3 - (\hat {\bf d}^{3}\cdot \boldgreek{ \sigma })\tau_{3}  \right)\Sigma_{0} (\bk
,\omega),\cr
\Delta ( {\bk ,\omega})&=& -
\tfrac{1}{4}\left (
(\hat {\bf d}^{1}+i \hat {\bf d}^{2})\cdot \boldgreek{\sigma } \right)\Sigma_{0} (\bk
,\omega).
\end{eqnarray}
$\Sigma_{N}$ describes a kind of odd-frequency
magnetism (with no onsite magnetic polarization).  The second-term
$\Delta (\bk ,\omega)$ in \eqref{comps}
describes a triplet gap function,
with a complex d-vector 
$\hat 
{\bf d}^{1}+ i \hat  {\bf d}^{2}$ which breaks time-reversal
symmetry. 

However, hidden from immediate view, is the fact that the Andreev
scattering $\Delta (\bk ,\omega)$ actually describes   a pair density wave. 
To see this, let us 
now transform our solution
back to the original electron gauge. Reversing the transformation
\eqref{Qvector}, we see that in the original gauge
$(V_1,V_{2},V_{3},V_{4})_{{\bf R}_{j}} = 
 \exp[i{\bf Q\cdot
R}_j] (1,-i,1,i)V_{0}$. 
Now the hyper-octagon lattice can be viewed as made of
four-atom coils marked by $1,2,3,4$ on Fig. 1, arranged on BCC
lattice.  From \eqref{Qvector} it follows that the  $\hat {\bf d}_{1,2}$
alternate along the coil and
between the center and corners of the BCC lattice. 
In other words, the
magnetic vector $\hat  {\bf d}^{3}$ is uniform, but the
superconducting d-vector  $\hat {\bf d}^{1}_{j}+i \hat {\bf d}^{2}_{j} 
$ is staggered between neighboring sites, forming a pair
density wave (PDW). 

 {\it The Goldstone modes and topology.} 
 We briefly touch on the topic of collective excitations and
 topology. As pointed out above, the gauge invariant quantities
 must connect two $V$ fields  at different points
by a string of gauge fields.  However, 
below the melting temperature $T_{c1}$  we can safely
 forget about Z$_2$ gauge fields and work in a fixed gauge $u_{(i,j)} =
 1$.  In this situation, the spinor $V$ acquires the status of a true
 order parameter. Symmetry dictates that {well below the transition where the amplitude fluctuations are weak} the Ginzburg-Landau free
 energy density $f$ for the low energy sector  is given by 
\bea
 && f [x] = \frac{\rho}{2}\Big|\Big(-i\vec\nabla +
{e} \vec A\Big)z_{\s}\Big|^2 
+
\frac{\vec{B}^{2}}{8\pi}
- g\mu_{B}\vec B \cdot ( z\dg \vec\s z) \cr &&\label{F1}
\eea
where $\vec{B}=\nabla\times \vec A$ and as before, $V_{j\sigma }=
({\rm V }/\sqrt{2}) z_{\sigma }$. We have also included a
Zeeman coupling. 
%This also can be recast as a version of Principal Chiral Field model:
%\bea
%&& f (x) =  \frac{\rho}{2}
%\mbox{Tr} \Big[\Big| (-i\vec\nabla+ {e}\vec{A}) U\Big|^{2}\Big] - g\vec B\mbox{Tr}( U^+\vec\s U) +\nonumber\\
%&& \frac{1}{8\pi}[\vec\nabla\times \vec A]^2,
%\label{F2}
%\eea

Since the spinor $z$ is defined  
by three Euler angles, transforming under the double-group $SU (2)$, 
small fluctuations of the order parameter consist of three Goldstone modes, one of them being higgsed if the condensate is charged. 
  The free energy  \eqref{F1} has been discussed in various contexts, in particular in connection with multi-band superconductivity when the superconducting pairing takes place on  Fermi surface consisting of multiple sheets \cite{faddeev,Babaev1}. The case of zero electric charge $e=0$ emerges in connection with frustrated magnetism \cite{Shifman,Batista}.  It is distinct from the conventional superconductivity due to the different topology of the order parameter manifold; here it is of the S$^3$ sphere.  %Novel topological excitations $\pi_{3} (SU (2))= \mathcal{Z} $.
Since $\pi_{1} (SU (2))=0$, 
it forms a fragile superconductor
with a Meissner effect, but \underline{zero} critical current. 
%It describes a fragile superconductor unable to support nonzero critical current, but displaying Meissner effect. 
The existence of the integer-valued topological invariant in three dimensions $\pi_{3} (SU (2))= \mathcal{Z} $
\begin{eqnarray}\label{}
  %Q = -\frac{1}{2\pi^2}\int \rd^3 x \Big(\vec\omega_x[\vec\omega_y\times \vec\omega_z]\Big), ~~
%\vec\omega_{\mu} &=& -\frac{\ri}{2}\mbox{Tr}[\vec\s U^{-1}\p_{\mu}
%U],\cr
Q
= \frac{i }{24\pi^2} \int \rd^3 x \epsilon_{\mu\nu\lambda}\mbox{Tr}
\Big(U^+\p_{\mu} U U^+\p_{\nu} U U^+\p_{\lambda} U\Big),\ \  \label{Q}
\end{eqnarray}
where 
\begin{equation}\label{}
 U = \left(
\begin{array}{cc} z_{\uparrow} & -z^{*}_{\downarrow}\\
z_{\downarrow} & z^*_{\uparrow}
\end{array}
\right).
\end{equation}is an SU(2) matrix,
suggests a possibility of nontrivial topological defects of the
kind found in the Skyrme model of nuclear matter (for a review,
see, for example, \cite{Skyrme}). The latter model, however, contains
a term with four derivatives whose presence is required to prevent
the Skyrmions from collapse. It has been argued by \cite{faddeev} and
later by \cite{Babaev1,Shifman} that such terms are generated once one
takes into account the fluctuating magnetic field. These authors 
suggested that in this case, the the Ginzburg Landau 
free energy (\ref{F1}) admits additional 
knotted
solitons, or Hopfion topological configurations
\cite{faddeev,Babaev1,Shifman}. However, 
numerical calculations performed in
\cite{Hopf} indicate that the Hopfions are unstable leaving their
existence an open question.

{\it Conclusions.} We have presented a model of a three
dimensional Kondo lattice which exhibits a remarkable range of 
properties associated with strong correlations.   Some of these 
properties, such as pair density wave formation,  have been
observed experimentally, others, like odd-frequency pairing and the
formation of a neutral Fermi surface, 
have been a matter of ongoing debate. The success of our approach is
based on the fact that we are able to treat the strong correlations
exactly in the asymptotic region of weak Kondo coupling.  
%\begin{itemize}

There are three key aspects to our work. First, the condensate represents a new class of superconductivity, with the fractionalized order parameter  that transforms
under a double group.  This feature leads to a number of robust
consequences. 
In particular, the group topology determines the fragility of the superconducting order: although it displays diamagnetism (Meissner effect), the critical current is zero. Since the first homotopy $\pi_1(SU(2)) =0$, there are no vortices, but a nontrivial third homotopy  $\pi_{3} (SU (2))= \mathcal{Z} $ suggests a possibility of such topological excitations as Hopfions or hedgehogs. 
The gauge invariant order parameter also 
breaks  time reversal symmetry in a fashion that is protected by
Kramers theorem and independent of crystal lattice, thus the
transition into this state will not split
under strain. 
Secondly, the proposed superconducting order forms a PDW that 
coexists with a novel form of magnetic order. 
This property is related to the
fact that the nested Fermi surfaces of the conduction electrons and
the Majoranas are centered at different points in the Brillouin
zone, so that when an electron enters the spin liquid, it needs to
borrow momentum from the condensate.
Thirdly,   the low temperature excitations are described by a quasi-neutral
Fermi surface whose existence is guaranteed by the mismatch between the
quantum numbers of electrons and the Majorana spin excitations,
and the long-range coherence of the 
charge $e$ condensate. 
This guarantees that even in the
situation of perfect nesting (half filled conduction band) there is a residual
Majorana Fermi surface. 

We end by noting that one of the key features of the current model is
the stabilization of an underlying $Z_{2}$ spin liquid inside a Kondo
lattice,  by orbital degrees of freedom which decouple by forming 
a kind of valence bond solid. 
This is a situation that conceivably, could occur in quantum materials,
such as the topological Kondo insulator SmB$_{6}$, which 
under field, exhibits Quantum oscillations
reminiscent of a bulk Fermi surface\cite{Sebastian_Science2015,Erten:2017bj}. This material 
is thought to involve a quartet spin-state interacting
with a conduction sea\cite{ABC2013,Legner15}.  It is interesting
to speculate that these orbital degrees of freedom may, under some
circumstances, freeze into a valence bond-solid, stabilizing
an underlying Majorana spin liquid within the Kondo insulator. 

%By
%combining these two facts we see that the direction of these vectors
%alternate between the sites of the hyper-octagon lattice. Meanwhile
%vector ${\bf d}_3$ does not alternate, allowing it 
%to directly  couple to the magnetic field. 
%To simplify
%matters we have considered only the case of perfectly nested Fermi
%surfaces which is achieved in our model at half-filled conduction
%band. We postpone the discussion of effects created by imperfect
%nesting (finite doping or next-to-nearest neighbor hopping) till
%future publications.

%${\bf Q} = (\pi,\pi,\pi)$. The primitive lattice vectors are on a BCC,
%defined by the $a_1=(1,0,0), a_2 = (1/2,1/2, -1/2)$ and $a_3 = (1/2,1/2,1/2)$.  While ${\bf Q.a}_1=\pi,  {\bf Q.a}_2 = \pi/2$ and ${\bf Q.a}_3= 3\pi/2$.

%\item Novel topological excitations $\pi_{3} (SU (2))= \mathcal{Z} $.
%No vortices $\pi_{1} (SU (2))=0$, so forms a fragile superconductor
%with Meissner effect, but zero critical current. 

%{\bf Old text.} Quasi-neutral Fermi surface.  Current operator is finite with
%next nearest neighbor hopping.  ``Ghost'' or ``transverse'' metal,
%with a finite transverse conductivity. 

\begin{acknowledgments}
\textit{Acknowledgments:}
This work was supported by Office of Basic Energy Sciences, Material
 Sciences and Engineering Division, U.S. Department of Energy (DOE)
 under Contracts No. DE-SC0012704 (AMT) and DE-FG02-99ER45790
 (PC and AP). PC is grateful for discussions with Premi Chandra, Tom Banks, Yashar
Komijani, Eduardo Fradkin and Subir Sachdev and the 
support of the Aspen Center for Physics under NSF Grant PHY-1607611,
where part of this work was completed. AMT is grateful to E. Babaev,
M. Shifman and D. Schubring for valuable discussions. 

%Sciences grant DE-FG02-99ER45790.  
\end{acknowledgments}
%\bibliography{GeneralHeavyFermion}
%\bibliographystyle{apsrev4-1}

% Please enter all references in GeneralHeavyFermion.bib file in
% bibtex format. 
%\bibliography{GeneralHeavyFermion.bib}


\end{document}

\title{Supplemental Material for ``A solvable 3D Kondo lattice exhibiting pair density wave, odd-frequency pairing and order fractionalization"
}
\author{Piers Coleman}
\affiliation{
Center for Materials Theory, Department of Physics and Astronomy,
Rutgers University, 136 Frelinghuysen Rd., Piscataway, NJ 08854-8019, USA}
\affiliation{Department of Physics, Royal Holloway, University
of London, Egham, Surrey TW20 0EX, UK.}
\author{Aaditya Panigrahi}
\affiliation{
	Center for Materials Theory, Department of Physics and Astronomy,
	Rutgers University, 136 Frelinghuysen Rd., Piscataway, NJ 08854-8019, USA}
\author{Alexei Tsvelik}
\affiliation{
Division of Condensed Matter Physics and Materials
Science, Brookhaven National Laboratory, Upton, NY 11973-5000, USA}
\date{\today}
\pacs{PACS TODO}
\begin{abstract}
{ 
}
\end{abstract}

\maketitle

\noindent {\bf 1. $T_{c}$ and the logarithmic divergence in the pair
susceptibility}\\

The  nesting of the Fermi surfaces between the band electrons and the
Majorana fermions generates logarithmic singularities in ladder type
diagrams. This nesting originates from the simple fact that the
electrons and the Majoranas live on the same lattice and have the
identical (up to a factor) tight binding Hamiltonians. We can
calculate the transition temperature $T_{c}$ to logarithmic accuracy
by computing the leading logarithmic divergence in the charge $e$ pair
susceptibility. 

We consider the 
Hamiltonian
\begin{equation}\label{}
H= H_{c}+H_{YL} + H_{K}.
\end{equation}
Below $T_{c1}\sim 0.12 K$, the local moments fractionalize $\vec{S}_{j}\rightarrow  -\frac{i}{2}\vec{\chi}_{j}\times \vec{\chi}_j
$, 
so at half-filling, using equations (3) and  (7) of the main
text
\begin{eqnarray}\label{}
H_{YL}&=&  \frac{K}{2}\sum_{\bk \in BZ}
\vec{ \chi}\ \dg _{\bk l}
h (\bk )_{lm}
\vec{\chi}_{\bk m },\cr
H_{c}&=& - t\sum_{\bk\in BZ }
c\dg _{\bk \sigma l }  h (\bk)_{lm}
c_{\bk \sigma m }, 
\end{eqnarray}
while $h (\bk )$ is determined by Eq.(4) of the main text,
\begin{equation}\label{}
h (\bk )=
\pmat{
0 & i &  ie^{-i\bk\cdot \ba_2} & ie^{-i\bk\cdot \ba_1 }
\\-i & 0 & -i & ie^{-i \bk\cdot\ba_3}
\\- ie^{i\bk\cdot \ba_2} & i & 0 & -i
\\-ie^{i\bk\cdot \ba_1 } &  -ie^{i \bk\cdot\ba_3} & i & 0
},
\end{equation}
The indices
$l,m =1,...4$ mark different sites in unit cell. 
The Kondo interaction, taken from equation (1) of the main text is 
\begin{eqnarray}\label{eq3}
H_{K}&=&
J\sum_{j}
(-\frac{i}{2}
\vec{\chi}_{j}\times \vec{\chi}_j
)\cdot (
c\dg_{j}\vec{\sigma }c_{j}
) \cr &=& 
- \frac{J}{4} \sum_{j} 
c\dg_{j\alpha} 
[\sigma^{a},\sigma^{b}]_{\alpha \beta }c_{j\beta }
\chi_{j}^{a}\chi_{j}^{b},
\end{eqnarray}
where we have substituted $\vec{S}_{j}= -\frac{i}{2}\vec{\chi}_{j}\times \vec{\chi}_j
$ and used the identity $[\sigma^{a},\sigma^{b}] = 2 i
\epsilon_{abc}\sigma^{c}$.

\begin{figure}[h]
	\begin{center}
	\includegraphics[scale=0.15]{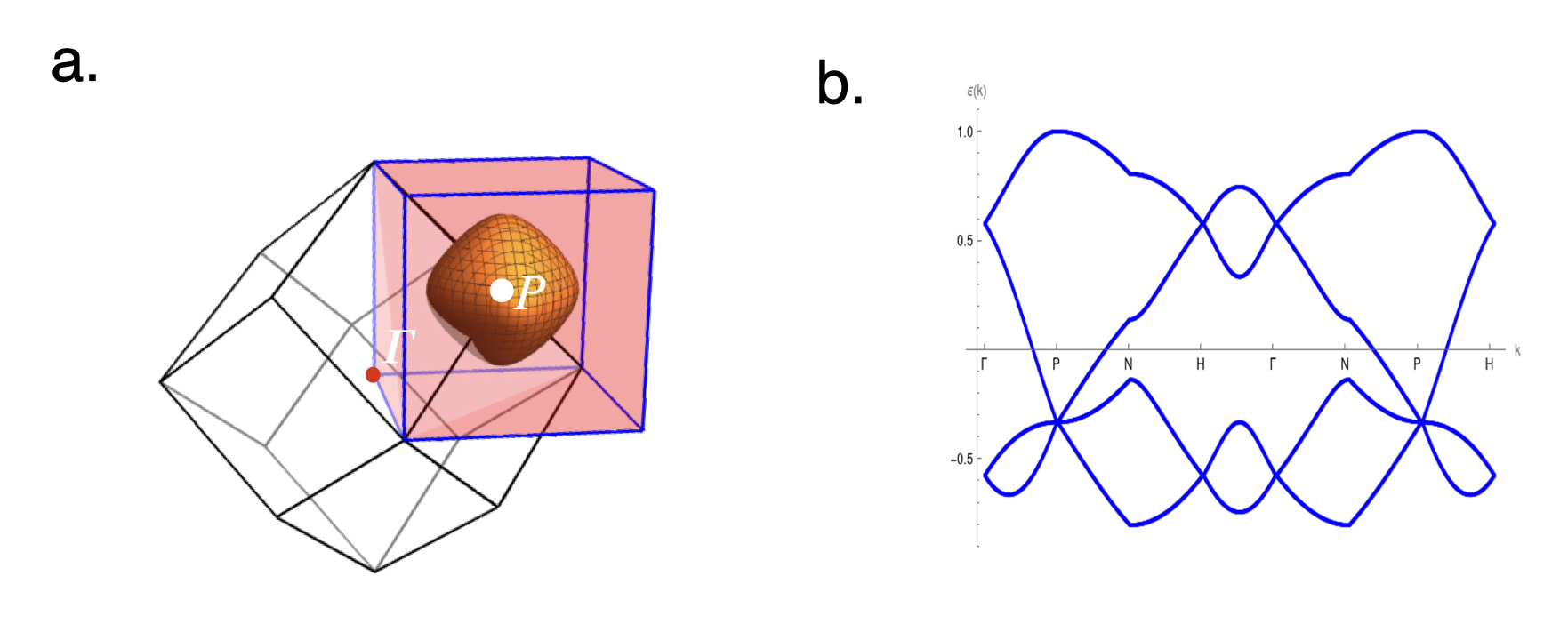}
	\caption{The Brillouin zone and the dispersion determined by
	Eq. 5 of the main text. \label{fig2}}
	\end{center}
\end{figure}

We now consider the self-consistent evaluation of the order parameter 
\begin{equation}\label{V}
V_{\alpha } (l) = \bigl\langle (\vec{\sigma }_{\alpha \beta }\cdot \vec\chi_{l})c_{l\beta }\bigr \rangle 
\end{equation}
to leading logarithmic accuracy in perturbation theory, using a
Feynman diagram expansion in $J$. 
We develop the Feynman diagram expansion in the standard way, starting
with a Wick expansion of $Z/Z_{0} =\langle T\exp [ - \int_{0}^{\beta
}H_{K} (\tau ) d\tau ]\rangle $, to obtain a
diagrammatic expansion for the Free energy in terms of linked cluster
Feynman diagrams. In the diagrammatic expansion, the bare propagators of conduction electrons and Majorana fermions 
are diagonal in the spin indices which we omit, and are given by 
\begin{equation}\label{}
\raisebox{-0.07truein}{
\epsfig{file=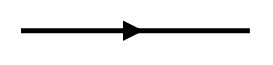,width =0.7in}} =
{G}^{c}_{lm} (k)= [ i\omega_{n}+t\hat  h (\bk )]^{-1}_{lm},
\end{equation}
for the conduction electrons and 
\begin{equation}\label{}
\raisebox{-0.05truein}{
\epsfig{file=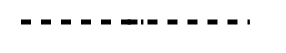,width =0.7in}} =
{G}^{M}_{lm} (k)= [ {i\omega_{n}- K \hat h (\bk )}]^{-1}_{lm},
\end{equation}
for Majorana fermions, where the $l,m = 1,2,3,4$ are the site indices
and we have also provided the Feynman diagram representation of the
propagators. Note the absence of an arrow on the Majorana propagator,
which means that when computing Feynman diagrams, one can assign an
arrow in either direction to any given Majorana propagator ($G_{M}
(k)= - ( G_{M} (-k)^{T}$). 
Diagonalizing the Hamiltonian $\hat h (\bk )_{jl}\psi_{\lambda }(l,\bk
)= \epsilon_{\lambda} (\bk )\psi_{\lambda} (j,\bk )$ we get 
\begin{eqnarray}
&& G^c(k)_{lm} = \sum_{\lambda =0}^3 \frac{\psi_{\lambda}(l,{\bf k}) \psi^*_{\lambda}(m,{\bf k})}{i\omega + t\epsilon_{\lambda}({\bf k})}, \nonumber\\
&& G^M(k)_{lm} = \sum_{\lambda =0}^3 \frac{\psi_{\lambda}(l,{\bf k}) \psi^*_{\lambda}(m,{\bf k})}{i\omega  -K\epsilon_{\lambda}({\bf k})}
\end{eqnarray}
where $\psi_{\lambda}(m,{\bf k})$ and $\epsilon_{\lambda}({\bf k})$ are eigenfunctions and eigenvalues of matrix $\hat h$. The resulting spectrum contains four bands;  only one band has a Fermi surface (see Fig. \ref{fig2}). Only this one contributes to the logarithmic divergence of the susceptibility. 
In the following calculation we truncate these Green's functions leaving only the contribution from $\lambda =0$ mode which crosses the chemical potential. 

The Feynman diagram expansion of
the Free energy involves the vertices
\begin{eqnarray}\label{}
\raisebox{-0.3truein}{
\epsfig{file=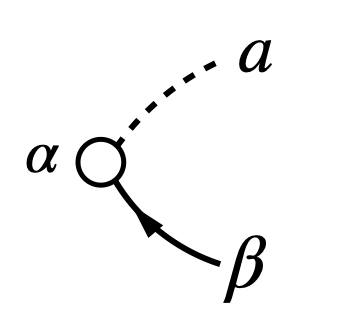,width =0.7in}} =\sigma^{a}_{\alpha \beta },
\end{eqnarray}
and 
\begin{equation}\label{}
\raisebox{-0.15truein}{
\epsfig{file=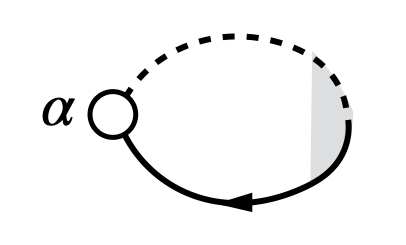,width =0.8in}} 
=\bigl\langle (\vec{\sigma }_{\alpha \beta }\cdot
\vec\chi_{l})c_{l\beta }\bigr \rangle = V_{\alpha } (l)
\end{equation}
represents the uniform anomalous order parameter at site $l$ (in the
gauge defined by equation (6) of the main text) while the
interaction vertex
\begin{equation}\label{}
\raisebox{-0.3truein}{
\epsfig{file=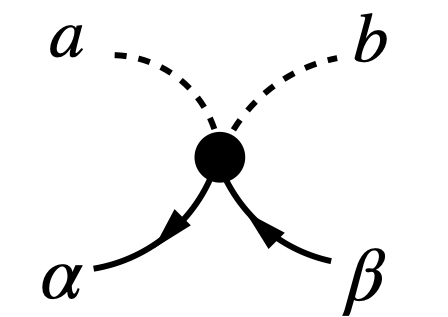,width =0.8in}} 
 = J\Lambda^{ab}_{\alpha \beta  }= \frac{J}{2 }
 [\sigma^{a},\sigma^{b}]_{\alpha \beta },
\end{equation}
where $\Lambda^{ab}_{\alpha \beta }$ can be written in the convenient
form
\begin{equation}\label{Lambda}
\Lambda^{ab}_{\alpha \beta } =
\frac{1}{2}[\sigma^{a},\sigma^{b}]_{\alpha \beta }= 
[\sigma^{a}\sigma^{b}]_{\alpha \beta }- \delta^{ab}\delta_{\alpha \beta}.
\end{equation}
The reduction of the denominator of \eqref{eq3} from $4$ to $2$ 
arises because there are two ways contract the Majorana fermions in
the interaction. 

To determine the transition temperature we 
derive a self consistent equation for the
pairing amplitude (\ref{V}), given by 
\begin{equation}\label{figA}
\raisebox{-0.3truein}{
\epsfig{file=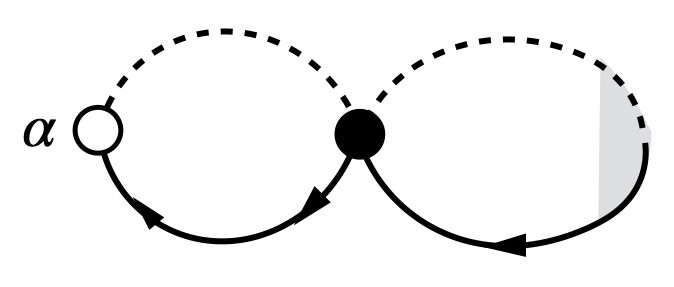,width =1.5in}} 
=\bigl\langle (\vec{\sigma }_{\alpha \beta }\cdot
\vec\chi_{l})c_{l\beta }\bigr \rangle = V_{\alpha } (l)
\end{equation}
which gives 
\begin{equation}\label{}
 V_{\alpha }(l) = J \sum_{m=1,4}\chi^{P}_{lm}V_{\alpha} (m)\label{BCS}
\end{equation}
where
\begin{equation}\label{}
\chi^{P}_{lm}
= -2 \frac{1}{\beta N_{s}}\sum_{\bk, i\omega_{n}} G^{c}_{lm} (k)G^{M}_{ml} (k)
\end{equation}
Note that the sum over momentum is a sum over the entire Brillouin
zone of the BCC lattice. 
For 
for convenience of calculation, we have used the particle-hole
symmetry of the Majorana propagator ($G_{lm} (k)= - G_{lm} (-k)$) 
to convert the ``pair''
susceptibility into a 
``particle-hole'', bubble (hence the $-1$).
The prefactor 2 in the above expression derives from
$\Lambda^{ab}_{\alpha \beta }\sigma^{b}_{\beta }= 2\sigma^{a}_{\alpha }$.
(We note that the factor $2$ replaces a naive factor of
3 obtained by the Hubbard-Stratonovich transformation, the difference 
corresponding to the distinction between a 
 Hartree and Hartree-Fock approximation.) We can divide the pair
 susceptibility into a divergent ``Cooper'' and a finite interband
susceptibility
\begin{equation}\label{}
\chi^{P}_{lm} = \chi^{C}_{lm} + \chi^{I}_{lm }
\end{equation}
 The divergent ``Cooper'' part of this
quantity dominates the susceptibility in the limit of low $J$, and
enables us to calculate the transition temperature to leading
logarithmic accuracy.  This component 
is obtained by projecting to the $\lambda=0$ band with a
Fermi surface, which gives
\begin{eqnarray}
\chi^{C}_{lm}&=& -2 \frac{1}{\beta N_{s}}\sum_{\bk, i\omega_{n}}
 \frac{
|\psi_0(l,{\bf k})|^{2}|\psi_0(m,{\bf k})|^{2}  
}{[i\omega_n +t \epsilon_0(\bk )][i\omega_n -K\epsilon_0(\bk )]}, \cr
&=&-2 T\sum_{i\omega_{n}}\int_{0<k_{l}<2\pi} \frac{d^3 k}{(2\pi)^3} \frac{
|\psi_0(l,{\bf k})|^{2}|\psi_0(m,{\bf k})|^{2}  
}{[i\omega_n +t \epsilon_0(\bk )][i\omega_n -K\epsilon_0(\bk )]}, \nonumber
\end{eqnarray}
Here, we have replaced $\frac{1}{N_{s}}\sum_{\bk } = 
\frac{1}{2}\int
\frac{d^{3}k}{(2\pi)^{3}} 
$, where the integral is over the Brillouin
zone of the BCC lattice (with lattice constant $a=1$, primitive unit
cell volume $a^{3}/2$ and Brillouin
zone volume $(2\pi)^{3}/ (a^{3}/2) =  2 (2\pi)^{3}
$ ).  We can replace
\[
\frac{1}{2}\int_{\bk \in BZ}
\frac{d^{3}k}{(2\pi)^{3}} 
=\int_{\bk \in 1/2 BZ}
\frac{d^{3}k}{(2\pi)^{3}} 
\]
where the integral is over the half-Brillouin zone defined by
$k_{l}\in [0,2\pi]$ ($l=1,3$). 

Now setting $V_{\alpha } (l) = V_{\alpha }$, then averaging over sites
we can write
\begin{equation}\label{avge}
 V_{\alpha } = J \chi_{1e}V_{\alpha}
\end{equation}
where
\begin{equation}\label{}
\chi_{1e}=\frac{1}{4}\sum_{m,l=1,4}\chi^{C}_{lm}.
\end{equation}
Now since $\sum_{m=1}^4|\psi_{0}(m,{\bf k})|^{2}=1$, we then obtain 
\begin{eqnarray}
\chi_{1e} &=&-\frac{1}{2}T\sum_{n}\int \frac{d^3 k}{(2\pi)^3}
\frac{1}{[i\omega_n +t \epsilon_0(\bk )][i\omega_n -K\epsilon_0(\bk
)]}\cr
&=&\frac{1}{4}
\int \frac{d^3 k}{(2\pi)^3}
\frac{{\rm th} \left(\frac{\beta t\epsilon_{0} (\bk )}{2} \right)+{\rm th}\left(\frac{\beta K\epsilon_{0} (\bk )}{2} \right)
}{(t+K) \epsilon_0(\bk )}
\end{eqnarray}
corresponding to equation (12) in the main text. 
We now replace
the momentum integral an energy integral over the density of states
writing
\begin{equation}\label{}
\chi_{1e} \approx \frac{1}{4}
\int_{-W}^{W} d\epsilon \rho (\epsilon)
\frac{{\rm th} \left(\frac{\beta \epsilon}{2} \right)+{\rm th}\left(\frac{\beta K\epsilon}{2t} \right)
}{(1+K/t) \epsilon},
\end{equation}
where 
\begin{equation}\label{}
\rho (E) = \frac{1}{2}\int_{ \bk \in BZ} \frac{d^{3}k}{ (2\pi)^{3}}\delta  (E+t\epsilon (\bk )),
\end{equation}
where the factor of $1/2$ arises because the volume of the Brillouin
zone is $2 (2\pi)^{3}$, and $W$ is the nominal band-width.
To obtain the logarithmically divergent part of the susceptibility,
we can approximate $\rho (E)$ by its value  (see section 2. eq (32) below)
\begin{equation}\label{}
\rho_{e} =\rho (0)=
\frac{2\sqrt{3}}{\pi^{2}t}, 
\end{equation}
which then gives
\begin{equation}\label{}
\chi_{1e} (T) = \frac{\rho_{e}}{(1+K/t)}\ln  \left(\frac{W}{T} \right)
\end{equation}
Thus, to logarithmic accuracy the critical temperature $T_{c2}$ is determined
by 
\begin{equation}\label{}
1=J \chi_{1e} (T_{c2})= \frac{\rho_{e} J}{1+K/t}\ln  \left(\frac{W}{T_{c2}} \right),
\end{equation}
The logarithmic divergence means that below
\begin{equation}\label{}
T_{c} = W \exp \left[-\frac{(1+K/t)}{J\rho_{e}} \right],
\end{equation}
corresponding to Eq. (12) of the main text. 
\\

\noindent{\bf 2. Calculation of the density of states}\\

Here we calculate the density of states of the conduction electrons 
and Majorana fermions at the Fermi energy. The energies
of the conduction electrons and Majorana fermions in the spin liquid
are $-t \epsilon (\bk )-\mu$ and $K\epsilon (\bk )$, where
$\epsilon=\epsilon (\bk )$ is an root of the quartic equation (5) in the main text
\begin{equation}\label{}
D (\epsilon)=\epsilon^4-6\epsilon^2 -8\epsilon (s_{x}s_{y}s_{z})
+ [4 (c_{x}^{2}+c_{y}^{2} + c_{z}^{2})-3]=0,
\end{equation}
where $c_{l} = \cos ( k_{l}/2)$ and $s_{l}= \sin ( k_{l}/2)$, ($l=1,3$).
We shall consider the case where $\mu=0$. The Fermi surface is
determined by the condition that $\epsilon (\bk ) =0$, or
\begin{equation}\label{}
c_{0}= \sqrt{c_{x}^{2}+ c_{y}^{2}+ c_{z}^{2}} = \frac{\sqrt{3}}{2}
\end{equation}
which describes an approximately spherical  
Fermi surface centered around $\bk = (\pi,\pi,\pi)$ of radius $|k|\sim
\pi/ (3a)$. 
We wish to
calculate the density of states at the Fermi energy 
\begin{eqnarray}\label{}
 \rho= 
\frac{1}{2} \int_{\bk  \in BZ}
\frac{d^{3}k}{(2\pi)^{3}}\delta (t \epsilon (\bk ))\nonumber
= \frac{1}{t}\int_{0<k_{l}<2\pi}
\frac{d^{3}k}{(2\pi)^{3}}\delta (\epsilon (\bk )),
\end{eqnarray}
where the integral is over the cube of side $2\pi$, $(0<k_{l }<2\pi)$,
$l= (1,2,3)$. Note that the volume of the Brillouin zone for the BCC lattice
is $2 (2\pi/a)^{3}$ , accounting for the normalization factor of $1/2$
in the momentum integral. We are able to reduce the integral to half
the Brillouin zone (the Majorana Brillouin zone) using the fact that
$\epsilon (\bk )= - \epsilon (-\bk )$.
It is convenient to change variables from momenta $\bk  = (k_{x},k_y, k_{z})$
to ${\bf c} = (c_{x},c_{y},c_{z})$, because in these co-ordinates, the Fermi
surface is a sphere of radius $\vert {\bf c} \vert = c_{0}=
\sqrt{3}/2$. Since  $dc_{l }= - \frac{1}{2}\sin (k_{l}/2)dk_{l}$ ($l =
1,3$), it follows that 
\begin{equation}\label{}
d^{3}k = \frac{8}{s_{x}s_{y}s_{z}} d^{3}c
\end{equation}
so that 
\begin{equation}\label{int1}
\rho = \frac{1}{t}
\int_{|c_{l}|<1} \frac{d^{3}c}{\pi^{3}s_{x}s_{y}s_{z}} \delta (\epsilon
({\bf c}))= \frac{1}{t}\int \frac{d\Omega }{\pi^{3}} \frac{c^{2}
dc}{s_{x}s_{y}s_{z}} \delta (\epsilon ({\bf c})),
\end{equation}
where we have switch to 3D polar co-ordinates. 
Now near the Fermi surface we can linearize $D (\epsilon)$
\begin{equation}\label{linearize}
D (\epsilon)= - 8 \epsilon s_{x}s_{y}s_{z} + 4 (c^{2}-c_{0}^{2}) + O (\epsilon^{2})
\end{equation}
so that for small $\epsilon$, 
\[
\epsilon = \frac{c^{2}-c_{0}^{2}}{2 s_{x}s_{y}s_{z}}
\]
i.e
\[
d\epsilon = \frac{c dc}{s_{x}s_{y}s_{z}},
\]
so  we can simply rewrite \eqref{int1} as 
\begin{equation}\label{}
\rho = \frac{c_{0}}{t}\int \frac{d\Omega}{\pi^{3}} d\epsilon \delta
(\epsilon) =  \frac{4 \pi c_{0}}{\pi^{3}t  } =
\frac{2\sqrt{3}}{\pi^{2}t }.
\end{equation}

In a similar fashion,  the density of states of the Majorana fermions in the 3D
Yao Lee model is given by
\begin{eqnarray}\label{}
 \rho_{M} 
= \frac{1}{2K}\int_{0<k_{l}<2\pi}
\frac{d^{3}k}{(2\pi)^{3}}\delta (\epsilon (\bk )) 
= \frac{2\sqrt{3}}{\pi^{2}K },
\end{eqnarray}
where the relative factor of $1/2$ appears because the initial
integral is restricted to half the Brillouin zone.

\begin{acknowledgments}
\textit{Acknowledgments:}
This work was supported by the Office of Basic Energy Sciences, Material
 Sciences and Engineering Division, U.S. Department of Energy (DOE)
 under Contracts No. DE-SC0012704 (AMT) and DE-FG02-99ER45790
 (PC and AP).

\end{acknowledgments}

\bibliographystyle{apsrev4-1}
\bibliography{trvb}